\documentclass[aps,prl,twocolumn,showpacs,psfig,superscriptaddress,longbibliography]{revtex4-2}

\usepackage{times}
\usepackage{graphicx}
\usepackage{float}
\usepackage{latexsym,amsmath,amssymb,bm,euscript,amsfonts}
\usepackage{color}
\usepackage{epstopdf}
\usepackage[colorlinks=true,linkcolor=blue,citecolor=blue]{hyperref}
\usepackage{soul}
\usepackage[normalem]{ulem}
\usepackage{mathrsfs}
\usepackage{lettrine}
\usepackage{bbding}
\usepackage{xspace}
\usepackage{textcomp}
\usepackage{textcase}
\usepackage{setspace}
\usepackage{chemformula}
\usepackage{siunitx}
\usepackage{physics}
\usepackage{xfrac}

\def\ECA{EuCo$_2$Al$_9$\xspace}
\def\BCA{BaCo$_2$Al$_9$\xspace}

\begin{document}
\title{Electrical Transport and Quantum Oscillations in the Metallic Spin Supersolid \ECA}

\author{Xitong Xu}
\thanks{These authors contributed equally to this work.}
\affiliation{Anhui Key Laboratory of Low-Energy Quantum Materials and Devices, High Magnetic Field Laboratory (CHMFL), Hefei Institutes of Physical Science, Chinese Academy of Sciences, Hefei, 230031, China}
\affiliation{Science Island Branch of Graduate School, University of Science and Technology of China, Hefei, Anhui 230026, China}

\author{$\hspace{-1.4mm}^{,\, \dag}$ Yonglai Liu}
\thanks{These authors contributed equally to this work.}
\affiliation{Anhui Key Laboratory of Low-Energy Quantum Materials and Devices, High Magnetic Field Laboratory (CHMFL), Hefei Institutes of Physical Science, Chinese Academy of Sciences, Hefei, 230031, China}

\author{Ning Xi}
\thanks{These authors contributed equally to this work.}
\affiliation{Institute of Theoretical Physics, Chinese Academy of Sciences, Beijing 100190, China}

\author{Mingfang Shu}
\affiliation{Anhui Key Laboratory of Low-Energy Quantum Materials and Devices, High Magnetic Field Laboratory (CHMFL), Hefei Institutes of Physical Science, Chinese Academy of Sciences, Hefei, 230031, China}
\affiliation{College of Sciences, China Jiliang University, Hangzhou 310018, China}
\affiliation{Key Laboratory of Artificial Structures and Quantum Control, School of Physics and Astronomy, Shanghai Jiao Tong University, Shanghai 200240, China}
\affiliation{Collaborative Innovation Center of Advanced Microstructures, Nanjing University, Nanjing 210093, Jiangsu, China}

\author{Haitian Zhao}
\affiliation{Anhui Key Laboratory of Low-Energy Quantum Materials and Devices, High Magnetic Field Laboratory (CHMFL), Hefei Institutes of Physical Science, Chinese Academy of Sciences, Hefei, 230031, China}
\affiliation{Science Island Branch of Graduate School, University of Science and Technology of China, Hefei, Anhui 230026, China}

\author{Jiajun Xie}
\affiliation{Anhui Key Laboratory of Low-Energy Quantum Materials and Devices, High Magnetic Field Laboratory (CHMFL), Hefei Institutes of Physical Science, Chinese Academy of Sciences, Hefei, 230031, China}
\affiliation{Science Island Branch of Graduate School, University of Science and Technology of China, Hefei, Anhui 230026, China}

\author{Guoliang Wu}
\affiliation{Institute of Theoretical Physics, Chinese Academy of Sciences, Beijing 100190, China}

\author{Hao Chen}
\affiliation{Lanzhou Center for Theoretical Physics, Key Laboratory of Theoretical Physics of Gansu Province, and Key Laboratory of Quantum Theory and Applications of MoE, Lanzhou University, Lanzhou, Gansu 730000, China}

\author{Miao He}
\affiliation{Anhui Key Laboratory of Low-Energy Quantum Materials and Devices, High Magnetic Field Laboratory (CHMFL), Hefei Institutes of Physical Science, Chinese Academy of Sciences, Hefei, 230031, China}

\author{Pengzhi Chen}
\affiliation{Anhui Key Laboratory of Low-Energy Quantum Materials and Devices, High Magnetic Field Laboratory (CHMFL), Hefei Institutes of Physical Science, Chinese Academy of Sciences, Hefei, 230031, China}
\affiliation{Science Island Branch of Graduate School, University of Science and Technology of China, Hefei, Anhui 230026, China}

\author{Ze Wang}
\affiliation{Anhui Key Laboratory of Low-Energy Quantum Materials and Devices, High Magnetic Field Laboratory (CHMFL), Hefei Institutes of Physical Science, Chinese Academy of Sciences, Hefei, 230031, China}

\author{Zhentao Wang}
\affiliation{Center for Correlated Matter and School of Physics, Zhejiang University, Hangzhou 310058, China}

\author{Chuanying Xi}
\affiliation{Anhui Key Laboratory of Low-Energy Quantum Materials and Devices, High Magnetic Field Laboratory (CHMFL), Hefei Institutes of Physical Science, Chinese Academy of Sciences, Hefei, 230031, China}

\author{Mingliang Tian}
\affiliation{Anhui Key Laboratory of Low-Energy Quantum Materials and Devices, High Magnetic Field Laboratory (CHMFL), Hefei Institutes of Physical Science, Chinese Academy of Sciences, Hefei, 230031, China}

\author{Haifeng Du}
\affiliation{Anhui Key Laboratory of Low-Energy Quantum Materials and Devices, High Magnetic Field Laboratory (CHMFL), Hefei Institutes of Physical Science, Chinese Academy of Sciences, Hefei, 230031, China}
\affiliation{Science Island Branch of Graduate School, University of Science and Technology of China, Hefei, Anhui 230026, China}

\author{Jie Ma}
\email{Corresponding authors: xuxitong@hmfl.ac.cn; jma3@sjtu.edu.cn; cx@lzu.edu.cn; w.li@itp.ac.cn; zhequ@hmfl.ac.cn}
\affiliation{Key Laboratory of Artificial Structures and Quantum Control, School of Physics and Astronomy, Shanghai Jiao Tong University, Shanghai 200240, China}
\affiliation{Collaborative Innovation Center of Advanced Microstructures, Nanjing University, Nanjing 210093, Jiangsu, China}

\author{Xi Chen}
\email{Corresponding authors: xuxitong@hmfl.ac.cn; jma3@sjtu.edu.cn; cx@lzu.edu.cn; w.li@itp.ac.cn; zhequ@hmfl.ac.cn}
\affiliation{Lanzhou Center for Theoretical Physics, Key Laboratory of Theoretical Physics of Gansu Province, and Key Laboratory of Quantum Theory and Applications of MoE, Lanzhou University, Lanzhou, Gansu 730000, China}

\author{Wei Li}
\email{Corresponding authors: xuxitong@hmfl.ac.cn; jma3@sjtu.edu.cn; cx@lzu.edu.cn; w.li@itp.ac.cn; zhequ@hmfl.ac.cn}
\affiliation{Institute of Theoretical Physics, Chinese Academy of Sciences, Beijing 100190, China}

\author{Zhe Qu}
\email{Corresponding authors: xuxitong@hmfl.ac.cn; jma3@sjtu.edu.cn; cx@lzu.edu.cn; w.li@itp.ac.cn; zhequ@hmfl.ac.cn}
\affiliation{Anhui Key Laboratory of Low-Energy Quantum Materials and Devices, High Magnetic Field Laboratory (CHMFL), Hefei Institutes of Physical Science, Chinese Academy of Sciences, Hefei, 230031, China}
\affiliation{Science Island Branch of Graduate School, University of Science and Technology of China, Hefei, Anhui 230026, China}

\date{\today}

\begin{abstract}
The discovery of spin supersolid and its giant magnetocaloric effect has opened a new arena in frustrated quantum magnets and cutting-edge cryogenics.
The intermetallic \ECA (ECA), for the first time, extends this intriguing phase from Mott insulators to a highly conductive metal~\cite{Nature2026}.
In this work, we systematically study the electrical transport properties of ECA, where itinerant electrons serve as a sensitive probe for the spin supersolid states.
We observe anomalies both in the temperature-dependent resistivity and field-dependent magnetoresistance and Hall signals, which are attributed to response of electrons to the Eu$^{2+}$ spins and their fluctuations.
Moreover, Shubnikov-de Haas quantum oscillations at high magnetic field reveal pronounced band splitting in the spin polarized state.
Our results reveal an intimate correspondence between electrical transport and magnetic transitions in ECA, deepening the understanding of this metallic spin supersolid.
\end{abstract}
\maketitle

\noindent
\textbf{Introduction.}
The spin supersolid --- an exotic magnetic state featuring the coexistence of long-range crystalline order and superfluid spin coherence --- has drawn intense attention due to the rich emergent quantum phenomena and its application potentials~\cite{Sengupta2007field, Sengupta2007chain,Chen56701, Gao2022QMats, LiusuoNBCP, Xiang2024Nature, Gao2024dynamics, Wang2023_SS, chen2024_KCSO,Zhu2024_KCSO,chi2024dynamical, t25p-x319, Sheng2025Continuum, huang2025,Xu2025,cui2025,Popescu2025,Ferreira_Carvalho_NBCP,2026-0074,LI2025}. 
It has been realized in several two-dimensional triangular-lattice Mott insulators such as Na$_2$BaCo(PO$_4$)$_2$~\cite{Gao2022QMats, Xiang2024Nature, Gao2024dynamics,Popescu2025,Ferreira_Carvalho_NBCP} and K$_2$Co(SeO$_3$)$_2$~\cite{Zhu2024_KCSO, chen2024_KCSO}, 
where significant low-temperature spin fluctuations and entropic effects lead to divergent Gr\"uneisen ratio at the quantum critical point, enabling efficient sub-Kelvin cooling via the adiabatic demagnetization refrigeration (ADR) technique~\cite{Xiang2024Nature}. 

\begin{figure*}[htbp]
\begin{center}
\includegraphics[clip, width=0.95\textwidth]{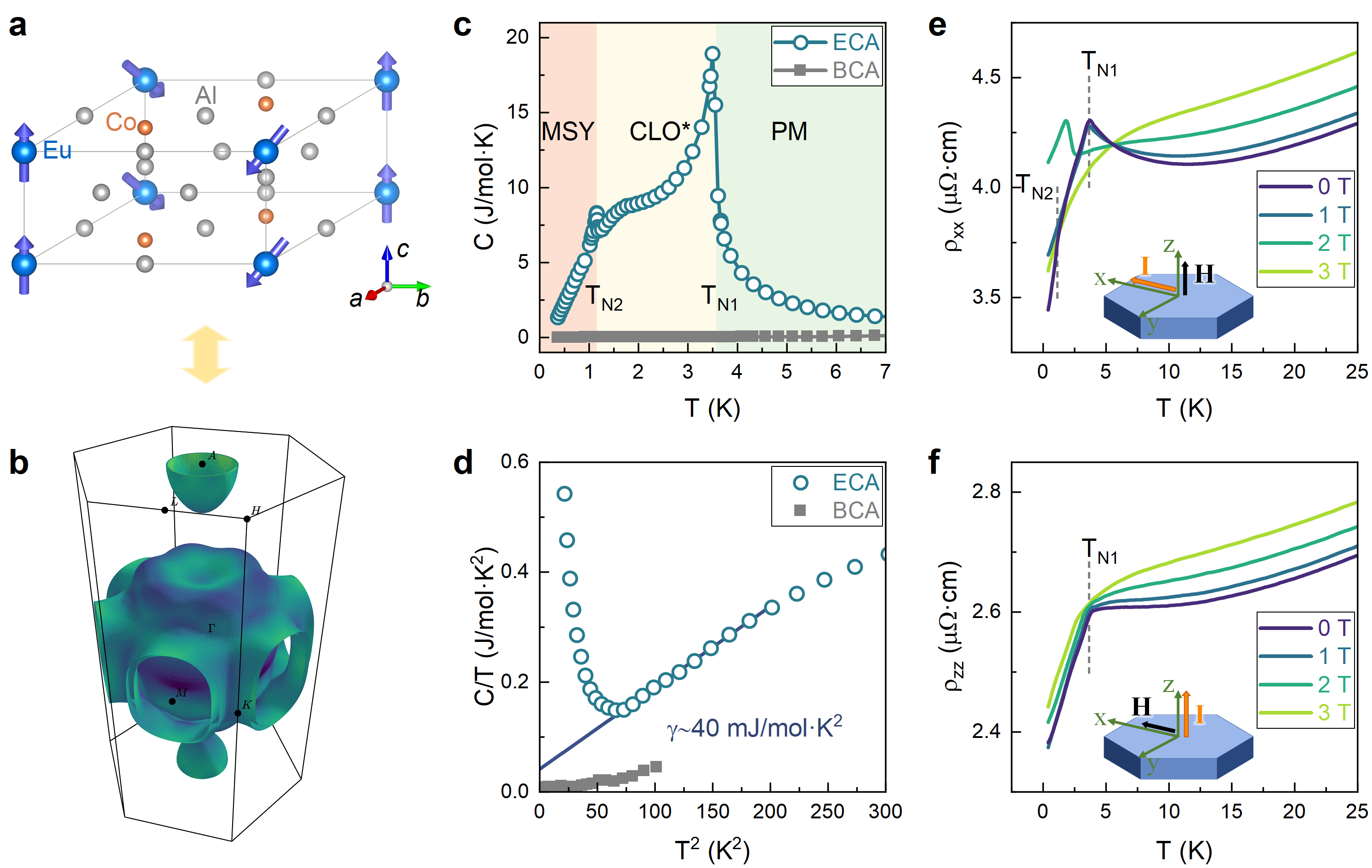}\\[1pt] 
\caption{Interplay between conduction electrons and supersolid spins in \ECA (ECA).
(a) Crystal structure of ECA, where the magnetic moments of Eu$^{2+}$ forms the metallic spin supersolid ``Y'' (MSY) state at low temperature. 
(b) Calculated Fermi surfaces of ECA in the paramagnetic (PM) state. 
(c) Specific heat of ECA at zero field.
The two peaks at $T_{N1}\simeq$ 3.5~K and $T_{N2}\simeq$ 1.1~K separate the high-temperature PM state, the fluctuating collinear ordered (CLO$^*$) state, and the MSY state.
The specific heat of the nonmagnetic counterpart \BCA (BCA) is also shown for comparison, which indicates negligible contributions of phonon and itinerant electrons in the temperature range of interest.
(d) Estimation of the electronic specific heat at 0~T using a linear fitting in the $C/T$ versus $T^2$ plot in the PM state.
The Sommerfeld coefficient is intercepted to be $\gamma \approx 40$~mJ/mol·K$^2$, suggesting a moderate yet important coupling between electrons and local moments.
(e) Longitudinal resistivity $\rho_{xx}$ for electric current $I$ along the $x$ direction (crystallographic $[11\bar20]$ direction). 
Magnetic field $H$ is applied along the $z$ direction ($c$-axis) from 0 to 3 T. 
The data show distinct features: a minimum above $T_{N1}$, a peak at $T_{N1}$, and a shoulder-like feature at $T_{N2}$.
Sketch illustrates the definition of the $xyz$ coordinates and the measurement configuration.
(f) Longitudinal resistivity $\rho_{zz}$ for $I$ along the $z$ direction and $H$ along $x$ direction.
There exists only one shoulder-like feature at $T_{N1}$.
}
\label{fig1}
\end{center}
\end{figure*}

The recent discovery of a metallic spin supersolid in the intermetallic compound \ECA (ECA) marks a significant advance~\cite{Nature2026}. 
This rare-earth intermetallic compound possesses a three-dimensinal, stacked triangular lattice arrangement of S=7/2 Eu$^{2+}$ ions as shown in Fig.~\ref{fig1}(a).
It has been established that the combination of Ruderman–Kittel–Kasuya–Yosida (RKKY) and dipolar interactions stabilizes a metallic spin supersolid ``Y'' (MSY) state as shown in Fig.~\ref{fig1}(a), where both out-of-plane three-sublattice order and in-plane spin superfluid order coexist~\cite{Nature2026,Xi2026}.
With increasing magnetic field applied along the $c$-axis, the system undergoes a transition into an up-up-down (UUD) 1/3 magnetization plateau phase at $\mu_0H_{c1}\sim0.3$~T, followed by a second supersolid phase --- the metallic spin-supersolid ``V'' (MSV) phase --- between $\mu_0H_{c2}\sim2.1$~T and $\mu_0H_{c3}\sim2.5$~T. 
Above $H_{c3}$, the system becomes a fully spin-polarized paramagnetic (PM) state.
Due to strongly fluctuating nature of the metallic supersolid states, ECA possesses a giant magnetocaloric effect that allows cooling to temperatures around 100~mK~\cite{Nature2026,Xiang2026}. 
Moreover, its ultrahigh electronic thermal conductivity, orders of magnitude larger than all known ADR coolants, overcomes heat-transfer limitations and establishes ECA as a superior solid-state refrigerant.

The coexistence of itinerant electrons and localized magnetic moments in ECA raises fundamental questions regarding how electrons mediate the spin supersolid order, and conversely, how electrons respond to magnetic transitions. 
In this work, we present a comprehensive electrical transport study of ECA, revealing distinct features of the interplay between magnetism and electrons through resistivity anomalies, magnetoresistance, Hall effect, and quantum oscillations. 
Our results establish that the electrical transport of ECA faithfully captures both the spin supersolid transitions and the electronic band features, advancing the understanding of this metallic spin supersolid state.

\bigskip
\noindent
\textbf{Electrical resistivity reflecting spin fluctuations.}
Single crystals of ECA and its nonmagnetic counterpart \BCA (BCA) were grown using the Al self flux method~\cite{Nature2026}.
Fig.~\ref{fig1}(b) shows the specific heat ($C$) of both ECA and BCA at zero field.
Compared with the vanishingly small specific heat of BCA, ECA possesses a large $C$ over a relatively wide temperature window due to the two step establishment of supersolid order at low temperatures.
It first undergoes a paramagnetic (PM) to a fluctuating collinear ordered (CLO$^*$) state at $T_{N1}\sim3.6$~K, through a phase transition with emergent U(1) symmetry~\cite{Nature2026,Xi2026}.
Similar fluctuating CLO$^*$ regime has also been observed as a floating Berezinskii–Kosterlitz–Thouless phase in 2D triangular-lattice compounds like Na$_2$BaCo(PO$_4$)$_2$~\cite{Xiang2024Nature, Gao2022QMats} and TmMgGaO$_4$~\cite{Li2020, Hu2020}.
With decreasing temperature, the superfluid order sets in at $T_{N2}\sim1.1$~K.
Besides the magnetic orders, the specific heat also provides information on the electronic state.
In the $C/T$ versus $T^2$ plot in Fig.~\ref{fig1}(d), the Sommerfeld coefficient is estimated to be $\gamma \approx 40$~mJ/mol·K$^2$ from a linear fitting in the PM state.
This value is larger than that of normal metal such as the constituent element Al~\cite{VanSciver2012} but much smaller compared to typical heavy-fermion systems~\cite{RevModPhys.56.755}, suggesting a moderate yet important coupling between electrons and local moments.

Fig.~\ref{fig1}(e) and (f) present the resistivities of ECA ($\rho_{xx}$ and $\rho_{zz}$) for current flowing within the $ab$-plane and along the $c$-axis, respectively.
Both $\rho_{xx}$ and $\rho_{zz}$ are exceptionally low (3.5 and 2.4~$\mu\Omega$·cm when approaching 0~K) compared with previously studied metallic coolants.
For instance, the resistivity of Yb-based Kondo-like metals is typically on the order of 100~$\mu\Omega$·cm at low temperatures~\cite{YSCZ2016,YPS2015,YCN2022,PhysRevB.109.024435,YNS2024,YNM2025}.
The resistivity data of ECA show an intriguing temperature dependence, with $\rho_{xx}$ clearly resolving both magnetic transitions at $T_{N1}$ and $T_{N2}$. 
Upon cooling, $\rho_{xx}$ exhibits a peak at $T_{N1}$, after which $\rho_{xx}$ decreases due to suppressed incoherent scattering. 
With the emergence of coherent in-plane spin order below $T_{N2}$, $\rho_{xx}$ decreases even faster.
On the other hand, $\rho_{zz}$ in Fig.~\ref{fig1}(f) displays only one shoulder-like feature at $T_{N1}$ and shows no discernible features at $T_{N2}$. 
The distinct resistivity behaviors indicate the spin fluctuations are prominent confined to the $ab$-plane, making them more detectable through in-plane electrical transport measurements than through out-of-plane measurements.
When an external field perpendicular to the current is applied, these anomalous features are suppressed and shifts towards lower temperatures.

\begin{figure*}[htbp]
\begin{center}
\includegraphics[clip, width=0.9\textwidth]{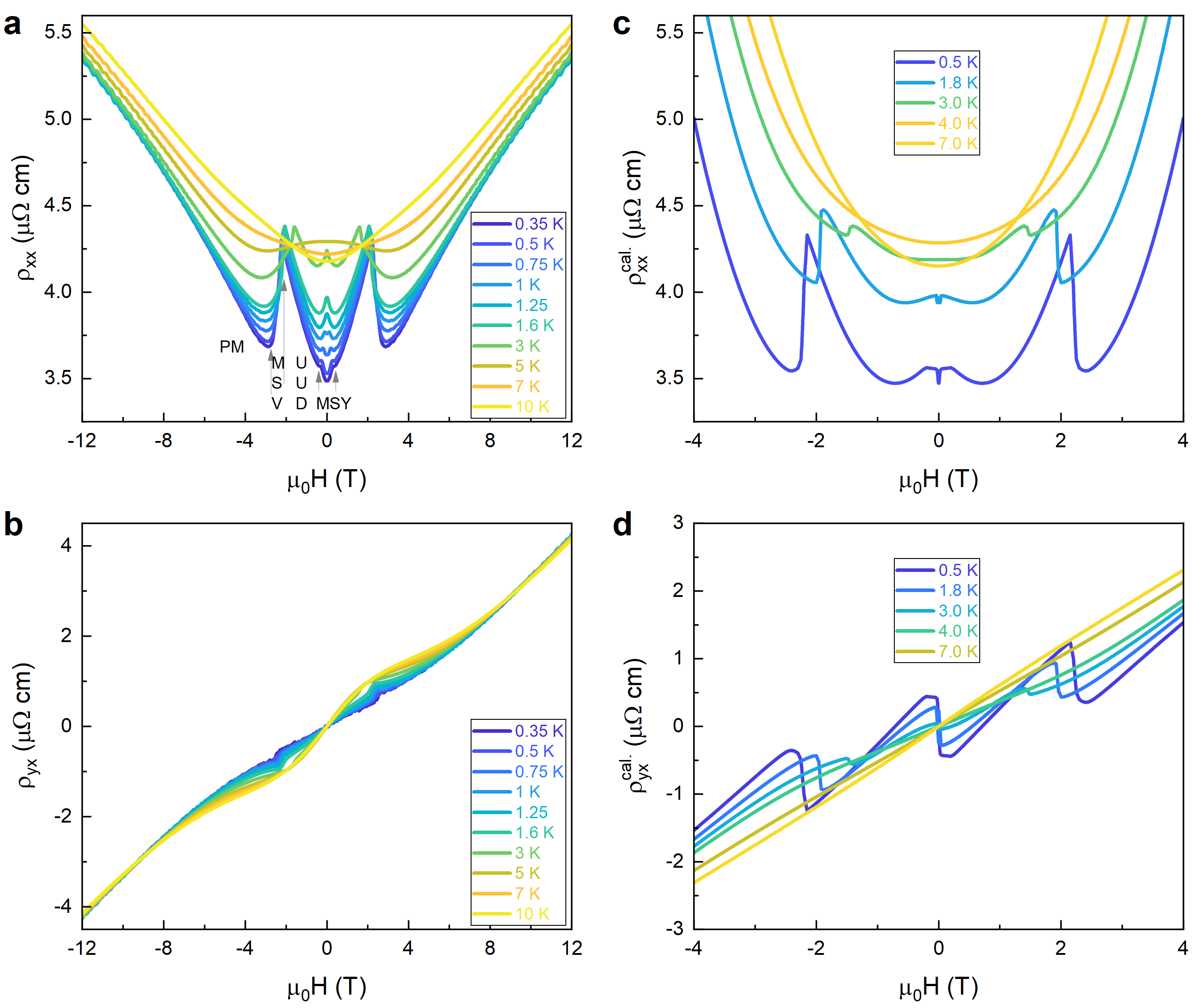}\\[1pt] 
\caption{Magneto-transport properties of ECA.
(a) Experimentally measured field dependence of $\rho_{xx}$ at representative temperatures, showing a dip in the MSY state and a steep decrease in the metallic spin supersolid ``V'' (MSV) state. 
Arrows denote the boundaries between MSY, UUD, MSV and PM phases at 0.35~K.
(b) Off-diagonal Hall resistivity $\rho_{yx}$, whose positive sign indicates dominant hole contribution.
(c) Simulated magnetoresistance, $\rho_{xx}^\mathrm{cal.}-\rho_0\propto(H-\alpha M)^2$, for $\lvert{\mu_0H}\rvert\leq$4~T, based on the experimental magnetization data in Ref.~\cite{Nature2026}. 
The parameter $\alpha$ denotes a mean-field coefficient fixed as 0.3 (see main text), which represents the coupling between electrons and local moments.
(d) Low-field zoom-in of simulated Hall resistivity, $\rho_{yx}^\mathrm{cal.}\propto(H-\alpha M)/(1+\eta H)$, where a slowly varying hole concentration with respect to field is assumed ($\eta=0.05$). 
}
\label{fig2}
\end{center}
\end{figure*}

\begin{figure*}[htbp]
\begin{center}
\includegraphics[clip, width=1\textwidth]{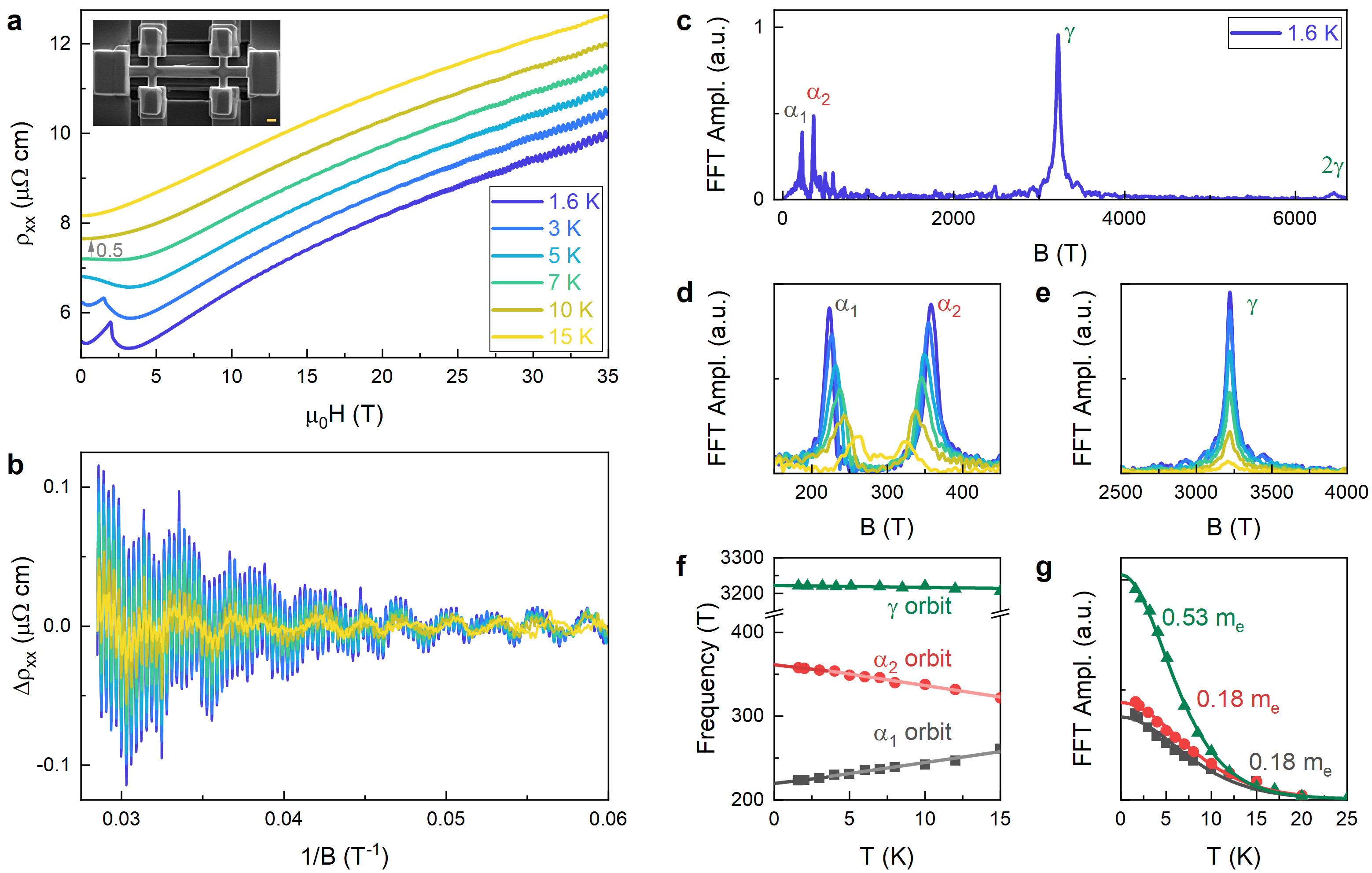}\\[1pt] 
\caption{Shubnikov-de Haas quantum oscillations (SdH QOs) in ECA.
(a) $\rho_{xx}$ for a FIB-milled Hall-bar sample of ECA up to 35~T, showing nonsaturating magnetoresistance and high-field SdH QOs.
The sample size is about $8\times1\times0.7$~$\mu m^3$ as shown in the inset, with field along the shortest direction ($c$-axis).
The curves are shifted vertically by 0.5~$\mu\Omega$·cm for clarity.
(b) Oscillatory part $\Delta\rho_{xx}$ of (a) after subtracting smooth backgrounds.
The magnetic induction intensity ($B=\mu_0H+\mu_0(1-N)M$) rather than $H$ has been adopted for the precise determination of the oscillation periods.
Here $M\sim0.39$~T for the saturation moment of ECA, and the demagnetizing factor $N$ is taken to be 0.5. 
(c) Fast Fourier transform (FFT) analysis of the QOs at 1.6~K in the 10-35~T window, revealing three major Fermi orbits ($\alpha_1$, $\alpha_2$, and $\gamma$).
(d) Temperature variations of the FFT spectra in the 5-14~T window.
(e) Temperature variations of the FFT spectra in the 10-35~T window.
(f) FFT peak positions of $\alpha_1$, $\alpha_2$, and $\gamma$ orbits with respect to temperature.
Solid lines represent linear fittings below and above $T_{N1}$.
(g) Fitting of the cyclotron masses of $\alpha_1$, $\alpha_2$, and $\gamma$ orbits from the peak values in (d) and (e), using the standard Lifshitz-Kosevich formula.
}
\label{fig3}
\end{center}
\end{figure*}

\begin{figure*}[htbp]
\begin{center}
\includegraphics[clip, width=1\textwidth]{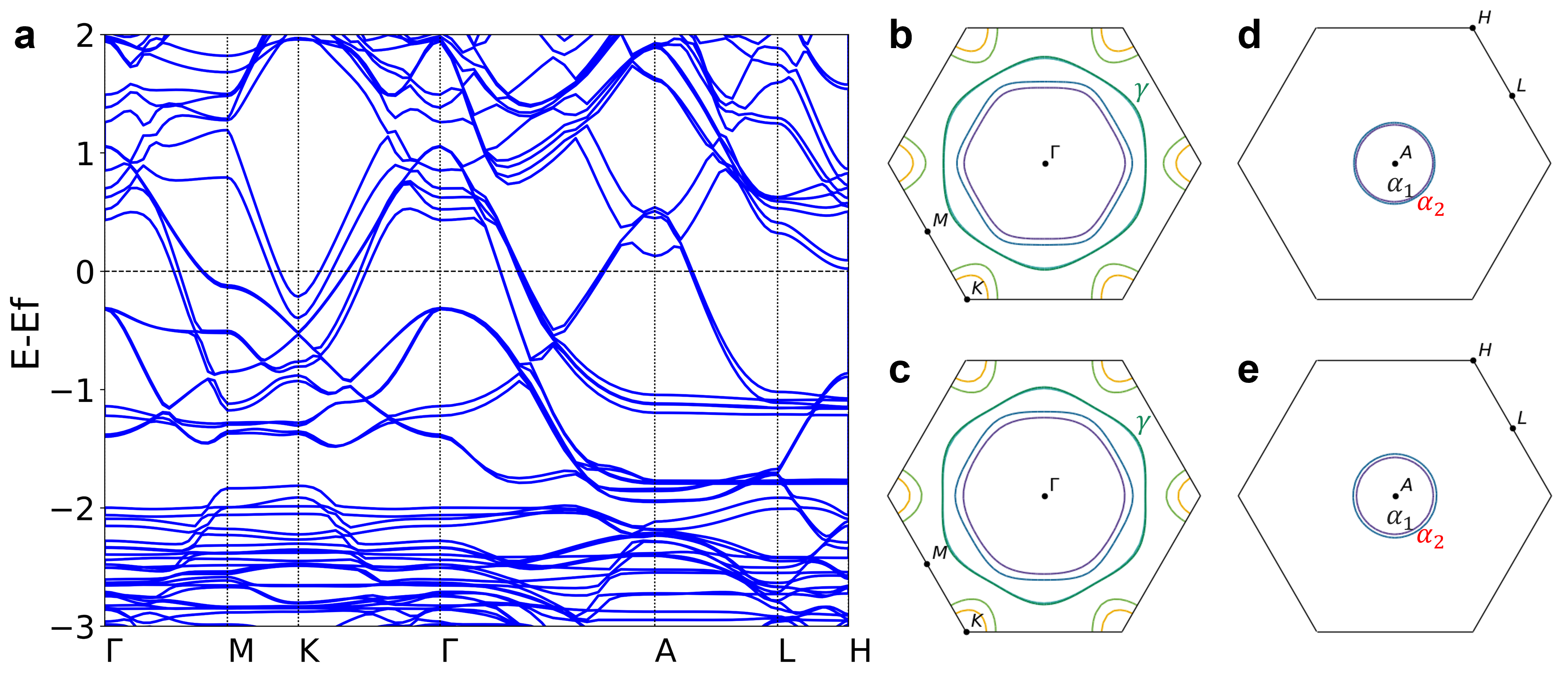}\\[1pt] 
\caption{Calculated electronic band structures of ECA.
(a) Band dispersions along high-symmetry lines, with all Eu$^{2+}$ spins fixed parallel to the $c$-direction.
(b) and (c) Fermi-surface cuts in the paramagnetic and polarized state, respectively, in the first Brillouin zone at $k_z=0$.
The largest hole pocket is labeled as the $\Gamma$ orbit.
(d) and (e) Fermi-surface cuts showing $\alpha_1$ and $\alpha_2$ orbits around $A$ point in the paramagnetic and polarized state, respectively, in the first Brillouin zone in the  $k_z=\pi/c$ plane.
}
\label{fig4}
\end{center}
\end{figure*}

\bigskip
\noindent
\textbf{Magnetoresistance and Hall effect under mean field coupling.}
The sensitive response of electrical transport to the magnetic phase transitions motivates us to study the magnetoresistance (MR) of ECA.
Two prominent features exist in the $\rho_{xx}(H)$ profiles shown in Fig.~\ref{fig2}(a): one is Shubnikov-de Haas quantum oscillations (SdH QOs) under high fields; the other is the multiple peaks and dips below the magnetic ordering temperature.
At 0.35~K, $\rho_{xx}$ displays a dip near zero field and a steep decrease between 2.1 and 2.5~T, corresponding to the nonclassical magnetization behaviors of the MSY and MSV states reported in Ref.~\cite{Nature2026}. 
In-between the MSY and MSV states, ECA is in the gapped UUD state, and the MR follows a normal, parabolic field dependence. 
The positive MR reappears above 2.5~T when all the moments in ECA are field-polarized.

To qualitatively explain these MR anomalies, we assume $\rho_{xx}^\mathrm{cal.}-\rho_0 \propto H_\mathrm{eff}^2=(H-\alpha M)^2$, where $H_\text{eff}$ is the effective field felt by conduction electrons, and $\alpha$ is a mean-field coefficient. 
This formula arises from the antiferromagnetic interaction between localized moments and itinerant electrons, which imposes an internal Weiss mean field on charge carriers. 
With experimentally obtained magnetization data from Ref.~\cite{Nature2026} and a fixed parameter $\alpha=0.3$, we simulate ${\rho_{xx}^{cal.}}$ in Fig.~\ref{fig2}(c). 
The theoretical calculation reproduces major experimental features in Fig.~\ref{fig2}(a), clarifying the origin of the unusual MR behavior in ECA. 
Interestingly, the mean-field treatment of $\rho_{xx}$ is inadequate in capturing the zero-field cusp observed between 1.6~K and 3~K in Fig.~\ref{fig2}(a). 
This discrepancy can be resolved by incorporating electron scattering from short-range-ordered magnetic moments in the CLO$^*$ regime~\cite{Wang2020}. 
Specifically, the fluctuating in-plane spin components contribute substantially to $\rho_{xx}$ between $T_{N1}$ and $T_{N2}$, but are rapidly suppressed upon the application of a magnetic field. 
Thus, the zero-field cusp marks the temperature regime where superfluidity emerges en route to a fully developed spin supersolid.

The Hall effect of ECA also manifests the mean-field coupling as shown in Fig.~\ref{fig2}(b).
To account for the overall slow variation of the positive slope, we assume a slowly varying hole concentration with respect to field, $\delta n(H)/n(0)=0.05H$.
Using the same extension as in the MR calculation, we set $\rho_{yx}^\mathrm{cal.}\propto(H-\alpha M)/n(H)$, which tracks the rapid changes of Hall effects across the UUD-MSV and MSV-PM phase transitions in Fig.~\ref{fig2}(d).
Nevertheless, such a simulation procedure exaggerates a zero-field switching behavior below 3~K which is hardly observable in experiments.
This discrepancy indicates the non-negligible role of quantum fluctuations at low temperatures in ECA.
It should also be noted that for simplicity, the change of relaxation time and multi-band effects were not considered, which may lead to an overestimation of the actual mean field.

\bigskip
\noindent
\textbf{Quantum oscillations and Fermi surface detection.}
In order to study the SdH QOs, we have fabricated a standard Hall-bar sample of ECA using focused ion beam (FIB) milling~\cite{bachmann2020manipulating}.
The sample size is about $8\times1\times0.7$~$\mu m^3$ as shown in the inset of Fig.~\ref{fig3}(a), with magnetic field applied along the shortest direction (the $c$-axis).
Its low-field magnetoresistance behavior resembles that of the bulk sample in Fig.~\ref{fig2}(a) except for a slightly larger residual resistance.
Under high fields up to 35~T, $\rho_{xx}$ remains unsaturated and follows quasi-linear profiles.

After subtracting a smooth background, the oscillatory component $\Delta\rho_{xx}$ is clearly resolved in Fig.~\ref{fig3}(b).
It is noted that the oscillations appear above 5~T, far above $H_{c3}$.
For precise determination of the oscillation periods, we use magnetic induction intensity ($B=\mu_0H+\mu_0(1-N)M$) rather than $H$, with $M\sim0.39$~T for the saturation moment of ECA and the demagnetizing factor $N$ taken to be 0.5. 
Fast Fourier transform (FFT) analysis at 1.6~K in Fig.~\ref{fig3}(c) reveals multiple oscillatory frequencies, the most prominent ones being 223, 358 and 3222~T, which are labeled as $\alpha_1$, $\alpha_2$ and $\gamma$ orbits, respectively.
According to the Onsager relation~\cite{shoenberg2009magnetic}, every QO frequency $F$ corresponds to an extremal cross-sectional area $A_k$
of the Fermi pockets in momentum space via $A_k=2\pi eF/\hbar$.
Assuming circular cross-sections ($A_k=\pi k_F^2$), the three orbits correspond to Fermi vectors $k_F$ of 0.08, 0.10 and 0.31 \AA$^{-1}$.

Interestingly, the peak positions of the $\alpha_1$ and $\alpha_2$ orbits exhibit strong temperature dependence as shown in Fig.~\ref{fig3}(d) and (f), shifting progressively towards each other as temperature increases, whose slopes appear to slightly change across $T_{N1}$.
At 15~K, the relative frequency change is up to 15\% (10\%) for the $\alpha_1$ ($\alpha_2$) orbit compared to data at 1.6~K.
In contrast, the $\gamma$ orbit remains unchanged over the temperature range considered.
The temperature variations of the oscillation frequency have also been reported in certain magnetic topological materials, including the Sb-doped MnBi$_2$Te$_4$~\cite{PhysRevB.103.205111}, PrAlSi~\cite{PhysRevB.102.085143}, RBi (R=Dy and Ho)~\cite{Nowakowska2023}, and Eu-containing EuMnBi$_2$~\cite{doi:10.1126/sciadv.1501117} and EuMnSb$_2$~\cite{PhysRevB.101.081104,PhysRevB.107.L081112}.
As discussed below, this phenomenon directly reflects a spin-polarization dependent electronic structure in ECA.

We also extract the cyclotron mass $m^\ast$ via the Lifshitz–Kosevich (LK) formula~\cite{shoenberg2009magnetic} as following,
$$R_T=\frac{2\pi^2 k_Bm^\ast T/eB\hbar}{\sinh 2\pi^2 k_Bm^\ast T/eB\hbar}$$
where $R_T$ represents temperature-dependent oscillation amplitude.
For the $\gamma$ orbit, $m^\ast$ is determined to be 0.53~$m_e$, whereas for the $\alpha_1$ and $\alpha_2$, $m^\ast$ is around 0.18~$m_e$.
These cyclotron masses are consistent with the moderate electronic specific heat in Fig.~\ref{fig1}(d).
The above parameters also enable the estimation of corresponding Fermi energy of orbits via $E_F=\hbar^2k_F^2/2m^\ast$ for parabolic bands, which is 700~meV, 140~meV and 230~meV for $\gamma$, $\alpha_1$ and $\alpha_2$, respectively.
These values are estimated with the assumption of parabolic dispersions, and would be modified if other band dispersion, like linear dispersion, is considered~\cite{yin2020discovery}.

\bigskip
\noindent
\textbf{DFT calculations and band structure.}
We have calculated the electronic structure of ECA in both the PM state where the direction of the spin of Eu$^{2+}$ is not constrained (three-dimensional Fermi surface in Fig.~\ref{fig1}(b)) and the spin polarized FM state where the spin of Eu$^{2+}$ is fixed to $z$ direction (band structure in Fig.~\ref{fig4}(a)) with the Vienna Ab initio Simulation Package~\cite{DFT1} using the Perdew-Burke-Ernzerhof (PBE) exchange-correlation functional. 
To account for the strong electronic correlations of the Co 3$d$ and Eu 4$f$ electrons, the PBE+U approximation was applied with U = 7.1 for Eu and U = 4.0 for Co. 

As shown in the band dispersions and Fermi surface cuts in Fig.~\ref{fig4}, ECA possesses multiple Fermi surfaces, including three sets of large hole pockets centered at the $\Gamma$ point, one pair of smaller hole pockets around the $A$ point, and two electron pockets around the $K$ points.
For the large Fermi surfaces around $\Gamma$ point, the Fermi velocity (the slope in Fig.~4(a) at $E_F$) of the innermost one is larger along $\Gamma-A$ line than along $\Gamma-K$ line, which might explain the difference in $\rho_{zz}$ and $\rho_{xx}$ in Fig.~\ref{fig1}(e-f). 
Using band parameters determined from QOs, the $\gamma$ orbit in the SdH QOs is determined to be the largest hole pocket around the $\Gamma$ point in Figs.~\ref{fig4}(b) and (c), whose $k_F$ shows no difference in the PM and FM states.
This orbit is in agreement with the angle-resolved photoemission spectroscopy measurement result reported in Ref.~\cite{Nature2026}, and plays the most prominent role in mediating the RKKY interactions.
On the other hand, the $\alpha_1$ and $\alpha_2$ orbits are expected to be the pair of hole pockets around the $A$ points.
Interestingly, these two pockets show significant splitting in the polarized spin state  compared to the PM state, as seen in Figs.~\ref{fig4}(d) and (e). 
This naturally explains the temperature-dependent QOs observed in Fig.~\ref{fig3}, where the thermal fluctuations disturb the polarized spin alignment and consequently affects the band splitting.
As for the different behaviors of the $\alpha_1$, $\alpha_2$ and $\gamma$ orbits, our DFT calculations have suggested the two $\alpha$ pockets are predominantly derived from Al orbitals, while the $\gamma$ pocket has significant contributions from Co orbitals~\cite{Nature2026}.
As the Eu$^{2+}$ ions are closely surrounded by Al$_6$ octahedra frameworks rather than Co atoms, the local environment may render that the $\alpha$ pockets more sensitive to the magnetic ordering of Eu spins.
Further theoretical studies will be needed to explicitly clarify the origin of the polarization dependent phenomenon.

\bigskip
\noindent
\textbf{Discussion.}
Unlike the spin supersolid phase in Mott insulators like Na$_2$BaCo(PO$_4$)$_2$~\cite{Gao2022QMats, Xiang2024Nature, Gao2024dynamics} and K$_2$Co(SeO$_3$)$_2$~\cite{Zhu2024_KCSO, chen2024_KCSO}, ECA can be recognized as an entirely new type of metallic spin supersolid with coupled local moments and electronic degrees of freedom.
This is, in part, visualized by our electrical transport measurements, where the supersolid spins not only govern fluctuation-dependent scattering processes but also provide a significant internal Weiss mean field.
The sensitivity of electrical transport to magnetic transitions is particularly useful, as it provides an accessible tool for probing supersolid transitions under high pressure, among other conditions.
These results establish ECA as a model system for studying metallic spin supersolidity as well as the cooperation between thermal fluctuations and the freedoms of electron and spin, and even the correlations between electrons and spins.
Constructing a full transport model explicitly integrating local moments, itinerant electrons, and their interplay, will be an important task in the study of metallic spin supersolidity.

Lastly, our combined SdH QO studies and DFT calculations suggest a dominating role of the largest hole pocket in the electrical transport. Since this hole pocket also controls the relative strengths of the RKKY interactions~\cite{Nature2026}, it is expected that electron or hole doping in the ECA system could effectively manipulate the metallic spin supersolid state, paving the way for further tunable, $^3$He-free ADR applications.


\bigskip
\noindent Note: After submission, we noticed a related study on \ECA proposing the existence of giant anomalous Hall effect~\cite{XU2026103285}.

\section*{Acknowledgments}
This work was supported by the National Key Research and Development Program of China (Grants No. 2024YFA1611101, 2024YFA1409200, 2024YFA1408303, 2023YFA1407300, 2022YFA1402704), 
the National Natural Science Foundation of China (Grants No. 12374129, 12534009, 12504186, 12534009, 12247101, 12374124), 
the Strategic Priority Research Program of Chinese Academy of Sciences (Grant No. XDB1270102), 
the CAS Project for Young Scientists in Basic Research (Grant No. YSBR-084), 
and the CAS Project (Grant No. JZHKYPT-2021-08). 
This work was also supported by Anhui Provincial Major S \& T Project (s202305a12020005),
Anhui Provincial Natural Science Foundation No.~2508085ZD013, 2408085J025. 
Y.L. is supported by the China Postdoctoral Science Foundation under Grant No. 2025M773379 and the Postdoctoral Fellowship Program of CPSF under Grant No. GZC20250072.
We thank the HPC-ITP for the technical support and generous allocation of CPU time. 
We acknowledge the support from Steady High Magnetic Field Facility (SHMFF, https://cstr.cn/31125.02.SHMFF). 
A portion of this work was supported by the High Magnetic Field Laboratory of Anhui Province under Contract No. AHHM-FX-2020-02.

\bibliography{ECA-bibliography}

\end{document}